\documentclass[sigconf, nonacm]{acmart}

\newcommand\vldbavailabilityurl{}
\newcommand\vldbpagestyle{plain} 

\usepackage{algorithmic}
\usepackage{graphicx}
\usepackage{textcomp}
\usepackage{xcolor}

\usepackage{comment}
\usepackage{mathtools}
\usepackage{amsmath}

\usepackage{calc}
\usepackage{color,soul}
\usepackage{balance}

\usepackage{algorithm}
\usepackage{algorithmic}
\usepackage{multicol}
\usepackage[toc,page]{appendix}
\usepackage{balance}
\usepackage{calc}
\begin{document}
\title{Scheduling of Intermittent Query Processing}

\author{Saranya Chandrasekaran}
\affiliation{%
  \institution{U R Rao Satellite Centre, ISRO}
  \institution{IIT Bombay}
  \postcode{400076}
}
\email{saranyac@cse.iitb.ac.in}

\author{S. Sudarshan}
\affiliation{%
  \institution{IIT Bombay}
  \streetaddress{}
  \city{}
  \state{}
  \postcode{400076}
}
\email{sudarsha@cse.iitb.ac.in}

\begin{abstract}
Stream processing is usually done either on a tuple-by-tuple basis or in micro-batches. There are many applications where tuples over a predefined duration/window must be processed within certain deadlines. Processing such queries using stream processing engines can be very inefficient since there is often a significant overhead per tuple or micro-batch. The cost of computation can be significantly reduced by using the wider window available for computation. In this work, we present scheduling schemes where the overhead cost is minimized while meeting the query deadline constraints. For such queries, since the result is needed only at the deadline, tuples can be processed in larger batches, instead of using micro-batches. We present scheduling schemes for single and multi query scenarios. The proposed scheduling algorithms have been implemented as a Custom Query Scheduler, on top of Apache Spark. Our performance study with TPC-H data, under single and multi query modes, shows orders of magnitude improvement as compared to naively using Spark streaming.
\end{abstract}

\maketitle

\pagestyle{\vldbpagestyle}
\begingroup\small\noindent\raggedright\textbf{}\\
\href{}{}
\endgroup
\begingroup
\renewcommand\thefootnote{}\footnote{
}\addtocounter{footnote}{-1}\endgroup

\ifdefempty{\vldbavailabilityurl}{}{
\vspace{.3cm}
\begingroup\small\noindent\raggedright\textbf{PVLDB Artifact Availability:}\\
The source code, data, and/or other artifacts have been made available at \url{\vldbavailabilityurl}.
\endgroup
}

\section{Introduction}

The main objective of stream processing is to process data as it arrives.  Many applications carry out analysis on data streams, and require results within a specified deadline, i.e. in real time. 
%For example in an online shopping portal, where a  number of offers are provided to customers, real time analysis of customer response can be done. Consider two input streams, one which contains the list of offers and another stream which contains the purchases made by the customers. These two streams can be merged to obtain relevant data such as the most purchased item and so on. This information can be used for deciding whether promotional offer for an item  has to be continued or not. 
%Another example for real time analytics is analysis of data from Stock market. By analysing this data, prospective shares which can be bought or sold can be suggested to the end user. 
Stream Processing Engines (SPEs) are widely used for doing such real time analytics. 
These systems are characterised by high input data rates and usually run a large number of concurrent queries. 
Stream Processing Engines usually do tuple-by-tuple processing or processing in micro batches, as tuples arrive.
However, doing computation eagerly is not needed for many applications.

For example, in an online shopping portal, one of the queries maybe to fetch  the total number of purchases done for each product in a day (or hour), and the result may be needed within some period after the end of the window on which the computation is being done. The application scenarios primarily include recurring queries. Our work was motivated by a leading E-Commerce site in India which requires analysis to be done on data collected over the day and the results must be made available at some time in the morning of the following day; they wished to move from daily to every few hours to have faster analysis. The same queries were run on successive windows, i.e. they are recurring queries; as per Zhang et al.\ \cite{ref_microsoft} 60\% of queries in Microsoft SCOPE are recurring in nature.  Also, Wang et al.\  \cite{ref_grosbeak} describe Grosbeak, a data warehouse implemented at Alibaba, to handle similar requirements where daily analysis queries must be processed within certain deadlines.

Stream processing engines such as Apache Spark and Apache Flink, update the aggregate as and when a new tuple arrives or when a micro batch of tuples arrive. This method of processing eagerly can lead to significant overheads.

For such queries with a longer deadline, as the results of the query are needed only at the deadline, tuples can be processed in larger batches. In the example considered, if the deadline of the query is say 2 hours after the end of the day, one option is to start computation after the end of the window.
However, in general, there may not be sufficient time to process the entire data in the gap between the end of the window and the query deadline.
In such cases, tuples may be collected for a duration of say one hour, and then processed together, and partial aggregates can be finally aggregated 
at the end. Thus, instead of eager processing, tuples can be processed in batches.  
This not only reduces the overall computation cost but also meets the deadline for delivering the query result.

The problem addressed in this paper is that of finding an optimal batch execution schedule that meets the required query deadlines while minimizing the cost. Here, cost refers to the total time required for processing the query. Our methods achieve resource savings in terms of reduction in the computation cost. Thus the overall CPU time requirement gets reduced compared to processing in micro batches.

We start with a scenario where only a single query needs to be scheduled
at a time and then consider scenarios with multiple queries where each can have its deadline. Using fewer batches reduces the overhead, so the goal is to minimize the cost while meeting the deadline.  Large window operations in streaming mode, often require data to be kept in memory for efficient processing.  There is thus a risk of encountering insufficient memory errors and the same was observed in our experiments presented in Section~\ref{results}.  Such memory issues are avoided in our processing scheme as processing is done intermittently on larger batches, allowing the use of algorithms that do not require the entire data to be memory resident.

Tang et al.\ \cite{ref_iqp} introduce the concept of Intermittent query processing, where parts of the query are executed on parts of the input at intermediate points, and the intermediate results are combined at the end to get the final result.  Shang et al.\ \cite{ref_crocodile_pap} propose query optimisation in such scenarios where query slackness is utilised to process (parts of) queries intermittently on data that have arrived till that point. This can reduce resource consumption in query processing.  Tang et al.\ \cite{ref_crocodile_demo} demonstrate the CrocodileDB system, which processes queries intermittently based on time or number of tuples, to reduce the overall computation cost. However, these works do not consider query deadlines.   Wang et al.\ \cite{ref_grosbeak} propose incremental computation over the available data. Though Wang et al.\ discuss scheduling, they do not provide further details on how the schedule is generated.  

In this work, we aim to optimise the scheduling of such Intermittent Queries by processing tuples in batches, in such a way that the overall computation cost is minimised, while satisfying the query deadline.  In general, stream processing engines run several queries simultaneously. The deadline requirements of one or more queries may coincide. Further, 
 queries may be added or deleted at any point in time. 
Our contributions are thus as follows:
\begin{enumerate}
    \item We first propose a method for scheduling batches of a single query, which guarantees to meet the deadline while minimizing the cost (assuming the deadline is feasible).
    \item We then consider the case of multiple queries, each with its deadline, which runs in a time-shared manner. The system may be dynamic where queries may be added at any time.  Here the goal is (a) to find batch sizes that keep the computation cost within some predefined bounds, and (b) to schedule the batches of each query based on its input availability, while prioritizing queries that have tighter deadlines.
    \item We consider both fixed and varying  input rates for tuple arrival.
    \item The proposed scheduling schemes have been implemented on a Custom Query Scheduler module, built on top of Apache Spark. 
    \item Experiments have been carried out under different scenarios on TPC-H data/queries, and demonstrate that our optimizations provide significant benefits in terms of reducing cost while meeting the query deadlines.
\end{enumerate}

%% What we have achieved ...

This paper is organised as follows.  Section~\ref{sec:probdef} gives the problem description and explains the factors that affect query scheduling.  Techniques for scheduling single queries are presented in Section~\ref{sec:singleq}, and 
techniques for scheduling multiple queries, under dynamic scenarios are described in Section~\ref{dynamic}.
Related work is presented in Section~\ref{relwork}.
Implementation aspects are described in
Section~\ref{sec_implementation} while experimental results are presented in Section~\ref{results}. Section~\ref{sec:concl} summarizes the conclusions and future work.

\section{Problem Description and System Definition} \label{sec:probdef}
In this section, we describe the problem specifications and the factors that impact the scheduling of queries with intermittent query processing. 

\subsection{Problem Description}

Let us consider an online shopping scenario. Input data arrives as a stream and queries typically analyse the data over a certain time duration and the output is expected within a deadline. The system is assumed to be a
soft real time system. Thus if the query results are delayed beyond the deadline, the utility of the results is not zero but decreases over time. Our techniques endeavour to complete query execution within the deadline provided it is feasible.  

Since the queries are recurring in nature, the time cost model can be derived from historical data. Initially, the scheduling strategies are explained for a single query case, in section \ref{sec:singleq}, where we assume that the input data rate can be modeled. Subsequently, in section \ref{dynamic}, we explain the scheduling techniques for a generic scenario considering  multiple queries along with uncertainties in input rates. We assume initially that each query runs on one input stream and can access multiple stored or static relations. Extensions to handle multiple streams involving joins are discussed as part of the implementation in Section \ref{sec_implementation}. 

%Most of the stream processing engines provide high level interfaces such as SQL for ease of use. 

Stream processing systems usually allow multiple queries to be processed simultaneously.  In this paper, we assume that queries are independent of each other. 
Also, the queries are assumed to support incremental operations \cite{ref_iqp}, and more specifically we assume that a query can be executed on parts of a stream, and the partial results can be combined later to get the final result.  Queries in streaming systems typically involve aggregation, and we assume that a final aggregation step is used to combine the partial aggregate results.

For example, consider a sample query where the total purchase done on the current day has to be  computed. Here the window ends at 12 AM, and the deadline can be defined at some time later, say 2:00 AM. Suppose the input stream, Order, consists of the following information: 

    \{ OrderID, ProductID, ProductName, Quantity, Price, CustomerID, Timestamp \}
    
The sample query to compute the total purchases done on the current day is given below.

\begin{center}
\label{query_simple_eg}
SELECT count(*) as TotalCount FROM orders \\ WHERE date =  TODAY
\end{center}

Here tuples of the query can be collected, say for every one hour, and then processed as a batch.  Only tuples that are available at the start of the execution of a given  batch are processed in that batch.  
Partial aggregates can be computed on each batch and the partial aggregates can be combined in a final  aggregation step at the end. 
While aggregation of partial aggregates can also be done intermittently to reduce the final aggregation cost, in this paper we restrict ourselves to strategies where partial aggregates are combined only in a single final aggregation step.

%The factors to be considered are discussed in Section \ref{sec_factors}. 

%Scheduling scheme for single query case under static scenarios, where the query requirements are known apriori is explained in Section \ref{static}. 
%Techniques for handling dynamic environment are presented in Section \ref{dynamic}.

\subsection{Factors for Query Scheduling} \label{sec_factors}

\begin{table}[t]
  \caption{Notation for Query Attributes}
  \label{tbl:query_attributes}
  \begin{tabular}{|l|l|}
    \toprule
    \textbf{Notation} & \textbf{Description}\\
    \midrule    
    queryID & \text{$Unique$ $Identifier$ $for$ $the$ $Query$} \\ \hline
    windStartTime & \vtop{\hbox{\strut \text{$Time$ $at$ $which$ $tuple$ $arrival$ $starts$}}}  \\  \hline 
    windEndTime & \vtop{\hbox{\strut \text{$Time$ $at$ $which$ $tuple$ $arrival$ $stops$}}} \\  \hline 
    $deadline_{Q}$ & \vtop{\hbox{\strut \text{$Time$ $by$ $which$ $the$ $Query$ $processing$}}  \hbox{\strut \text{  $must$ $be$ $completed$}}}
    \\ \hline 
    inputStream & \vtop{\hbox{\strut \text{$Denotes$ $the$ $query$ $input$ $stream$}}} \\ \hline 
    $inputRate_{S}$ & \vtop{\hbox{\strut \text{$Rate$ $at$ $which$ $tuples$ $arrive$ $for$}}  \hbox{\strut \text{ $inputStream$ }}} \\ \hline 
    numTupleTotal & \vtop{\hbox{\strut \text{$Total$ $number$ $of$ $tuples$ $to$ $be$ $processed$}}}  \\ \hline 
    minCompCost & \vtop{\hbox{\strut \text{$Time$ $required$ $for$ $processing$ $all$ $the$ $tuples$}} \hbox{\strut \text{$as$ $a$ $single$ $batch$}}} \\ \hline 
    slackTime & \vtop{\hbox{\strut \textbf{$The$ $maximum$ $time$  $beyond$  $which$  $the$ }}  \hbox{\strut \textbf{$query$ $processing$ $cannot$ $be$ $further$ $delayed$  }} \hbox{\strut \textbf{$without$ $missing$ $the$ $deadline$}} }\\
    \bottomrule
  \end{tabular}
\end{table}

The parameters of a query that affect scheduling decisions and relevant notations are given in Table \ref{tbl:query_attributes}. The main decisions to be made as part of query scheduling are the batch sizes and the time at which these batches must be scheduled, with the goal of minimizing the overall computation cost, while satisfying the deadline constraints.
Computation costs can be minimized by processing all the tuples of a given query in a single batch, but doing so may result in missing deadlines. The time cost model may be Linear or Non Linear. A simple linear cost model comprises  two factors: per tuple processing cost, and the per-batch overhead cost. 
The computation cost under the linear cost model is given in (\ref{eqn_cost_comp}). %Equation\ref{eqn_cost_comp}.

% If all tuples are processed together in a single batch, then the cost can be computed as in Equation\ref{eqn_min_cost}.

\begin{figure}[tb]
\centering
\fcolorbox{black}{white}{
\def\svgwidth{0.48\textwidth}
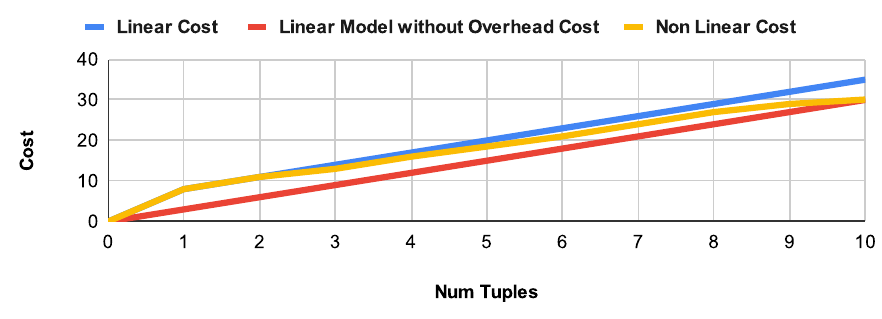}
\caption{Cost Model}
\Description{Cost Model}
\label{fig:cost_model_comparison}
\end{figure}

%\begin{equation}
%\begin{split}
%    minCompCost = (numTupleTotal *  tupleProcCost) \\ +  overheadCost \label{eqn_min_cost}
%\end{split}
%\end{equation}

\begin{equation}
\begin{split}
    compCost = (numTupleTotal *  tupleProcCost) + \\ (numBatches * overheadCost) \label{eqn_cost_comp}
\end{split}
\end{equation}

%Figure \ref{fig:tuplebytupleVsbatchedTuple} shows the comparison between doing tuple-by-tuple processing versus processing in batches. The top and bottom portions of Figure \ref{fig:tuplebytupleVsbatchedTuple} show tuple-by-tuple and processing in batches respectively and the cost incurred by these approaches.

Figure \ref{fig:cost_model_comparison}, illustrates different cost models. A sample linear cost model with a tuple processing cost of 3 and a constant overhead cost of 5 is shown as a shifted linear curve in Figure \ref{fig:cost_model_comparison}. A linear model without overhead cost and a non-linear model is also shown in  Figure \ref{fig:cost_model_comparison}. In general, the non-linear cost is expected to grow in a sublinear fashion. For our experiments, we use TPC-H queries with a piecewise linear model, as these queries fit well with the same. This is shown in section \ref{sec_cost_modelling}.

\section {Single Query under Static Scenarios: Requirements known Apriori} 
\label{sec:singleq}

In this section, we consider a static scenario where there is prior knowledge about the queries and their parameters, and we address the case where there is a single query with a deadline.  Dynamic scenarios with multiple queries are discussed later.
% or a number of queries wherein, only one of the queries has a deadline constraint to be met. 
Given a single query along with its parameters, the objective of the scheduling algorithm is to find the batch sizes such that the deadline is met while incurring the least computation cost. Our scheduling scheme is explained below. 

\subsection{Scheduling scheme} \label{static}

The Minimum Computation Cost, minCompCost, required to process all the tuples can be computed from the cost model for the query. Using this minCompCost,
The slack time of the query at the Window End can be computed as in (\ref{eqn_slack_time_comp}). %Equation \ref{eqn_slack_time_comp}.  

\begin{equation}
\label{eqn_slack_time_comp}
    slackTime  = deadline_{Q} -  windEndTime   - minCompCost     
\end{equation}

If there is sufficient duration to compute the query after the Window End Time by collecting all tuples together, then the query can be scheduled for processing starting at schStartTime as per (\ref{eqn_sch_st_time_1}). %Equation  \ref{eqn_sch_st_time_1}.

\begin{equation} \label{eqn_sch_st_time_1}
    schStartTime =  deadline_{Q}-minCompCost 
\end{equation}

If the slack time is negative, the query processing cannot be delayed until all tuples arrive. The query has to be processed in multiple batches starting prior to Window End Time. 

\begin{figure}[tb]
\centering
\fcolorbox{black}{white}{
\def\svgwidth{0.48\textwidth}
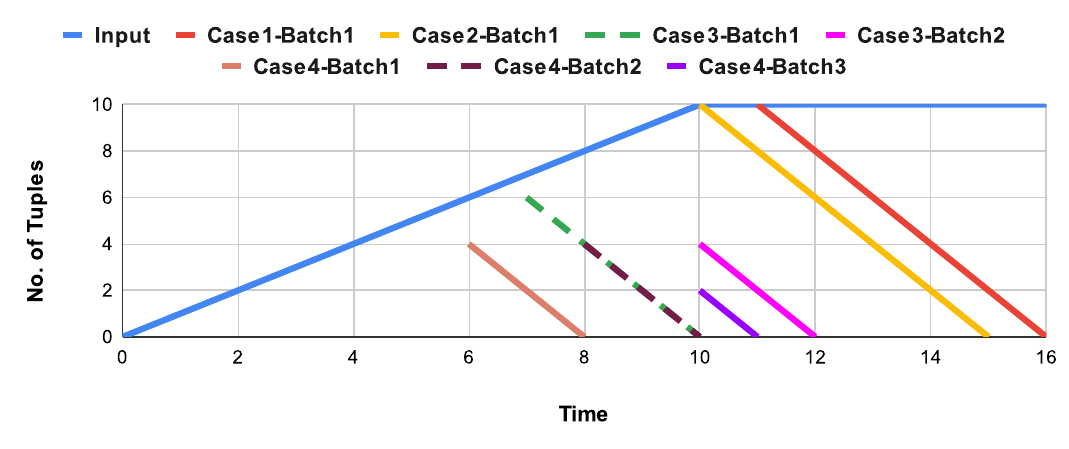}
\caption{Scheduling of Query for Case-1 to 4}
\Description{Scheduling of Query for Case-1 to 4}
\label{fig:query-sch-all-cases}
\end{figure}

Consider a query with an input rate of 1 tuple per sec, whose window start time and window end time be 1 and 10 respectively. Here, for the sake of simplicity, a linear cost model without overhead cost is assumed,  where 2 tuples can be processed per time unit. Let the final aggregation cost be negligible.  Now let us consider 4 cases with different deadlines of 16, 15, 12 and 11. Case-1 has positive slack time and can be processed after Window End at time 11. Processing of Case-2 has to be started at Window End, i.e. at time 10, to meet the deadline. The same is shown in Figure \ref{fig:query-sch-all-cases} as Case1-Batch1 and Case2-Batch1. 

Processing for Case-3 and Case-4 has to be done in batches. For Case-3, only 4 tuples can be processed from Window End Time to Deadline. So, the pending 6 tuples must be processed before Window End. These 6 tuples are available from time 6. The new deadline for completing the processing of these 6 tuples is Window End, i.e time 10. It will take 3 time units for processing these 6 tuples. So the processing can be scheduled at time 7,  where sufficient tuples are available and it also ensures that deadlines are met. The scheduling is shown in Figure \ref{fig:query-sch-all-cases} as Case3-Batch1 and Case3-Batch2.

For Case-4, only 2 tuples can be processed in the duration from  Window End Time to the deadline. So, the remaining 8 tuples must be processed before the Window End Time. The deadline for processing these 8 tuples is Window End Time. Eight tuples will be available by Time 8. From Time 8 to 10, only 4 tuples can be processed as per the cost model. The remaining 4 tuples must be processed before Time 8. So, the deadline for processing these pending 4 tuples is Time 8. Time taken for processing these 4 tuples is 2 time units. Thus, from Time 6 to 8, 4 tuples can be processed. The scheduling is shown in Figure \ref{fig:query-sch-all-cases} as Case4-Batch1, Case-Batch2 and Case4-Batch3. To summarize, the last batch of tuples is processed in the duration available from Window End Time till Deadline. The remaining tuples have to be processed before the Window End in one or more batches based on input availability.

\floatname{algorithm}{Function}
\begin{algorithm}
\caption{ScheduleWithoutAggCost (Query q) \label{algo_sch_single_slots}}
\begin{algorithmic}
\STATE \textbf{Input:} $Query$ $q$
\STATE \textbf{Output:} $schTuples[ ]$
\STATE $index$ $=$ $0$
\STATE $numTupleProc$ $=$ $0$
\STATE $timeDurLastBatch = q.deadline_{lastBatch}-q.windEndTime$
\STATE $numTupleProcLastBatch =$ \\
\STATE \hskip 2em $EstimateTuplesProcessed(q.queryID,timeDurLastBatch)$
\STATE $schTuples[index]=numTupleProcLastBatch$
\STATE $q.numTuplePending=q.numTupleTotal$
\STATE \hskip 3em $-numTupleProcLastBatch$
\STATE $timePt=q.deadline_{lastBatch}-timeDurLastBatch$
\WHILE{$(q.numTuplePending > 0)$}  
\STATE $ipAvailTime=InputTime(q.inputStream,$ \\
        \hskip10em $q.numTuplePending)$
\STATE $timeDur=timePt-ipAvailTime$
\STATE $numTupleProc=EstimateTuplesProcessed(q.queryID,$\\
 \hskip15em $timeDur)$
\IF{$(numTupleProc>q.numTuplePending)$}
\STATE $numTupleProc=q.numTuplePending$
\STATE $timeDur=Cost(q.queryID,q.numTuplePending)$
\ENDIF
\STATE $schTuples[++index]=numTupleProc$
\STATE $q.numTuplePending$ $-=$ $numTupleProc$
\STATE $timePt$ $-=$ $timeDur$
\ENDWHILE
\STATE \textbf{return $(schTuples)$ }
\STATE \textbf{End}
\end{algorithmic}
\end{algorithm}

The intuition behind our scheduling algorithm is as follows. If the query has non-negative slack time, it is scheduled in a single batch as in cases 1 and 2. For cases where it has a negative slack time, the query is processed in batches. Once all the batches are processed the final aggregation has to be done. Thus the deadline for the last batch 
$deadline_{lastBatch}$ has to account for the cost of the final aggregation, which we denote by $minCompCost_{Agg}$.
\begin{equation} \label{eqn_query_deadline}
    deadline_{lastBatch} =  deadline_{Q} - minCompCost_{Agg}
\end{equation}
% If the final aggregation cost is fixed, then the batch sizes can be determined by rescheduling the deadline for the  for the fixed aggregation cost.
The final aggregation cost however varies depending on the number and size of the batches.  
We therefore start with a final aggregation cost assuming two batches each containing all the groups, and compute $deadline_{lastBatch}$.
%We use a simplified model where the cost depends only on the number of batches.  Hence to start with, we assume that all tuples can be processed in 2 batches and reschedule the deadline to accommodate the final aggregation cost for 2 batches. With this modified deadline, 
Function-\ref{algo_sch_single_slots} \textit{ScheduleWithoutAggCost}, is invoked to determine the batch sizes that will incur the least computation cost while meeting the deadline $deadline_{lastBatch}$.
If the cost of final aggregation based on the batches determined by the function \textit{ScheduleWithoutAggCost} exceeds the assumption made earlier, then the deadline is modified considering a final aggregation cost with a larger number of batches each containing all the groups. This process is repeated by increasing the number of batches, until the final aggregation cost for the batches determined by the function \textit{ScheduleWithoutAggCost} is less than or equal to the assumed cost.

Function-\ref{algo_sch_single_slots} \textit{ScheduleWithoutAggCost} makes use of several subsidiary functions including:
\textit{InputTime(inputStream, numTuples)} which returns the time at which numTuples for the given input stream have been received, \textit{Cost(queryID, numTuples)} which returns the cost for processing numTuples for the given query, and  \textit{EstimateTuplesProcessed(queryID, duration)} which returns the number of tuples that can be processed in the specified duration for the given query.

Function-\ref{algo_sch_single_slots} \textit{ScheduleWithoutAggCost}, first computes the number of tuples that can be processed within the duration between the window end time and the deadline, $deadline_{lastBatch}$. Let this be denoted as $numTupleProcLast\-Batch$ and it represents the tuples that are processed in the last, i.e.\ the $n$\textsuperscript{th} batch. The pending tuples, which denote the difference between the total number of tuples and $numTupleProcLastBatch$, have to be processed within the window end time. Thus the deadline for processing the $n-1$\textsuperscript{th} batch has to be rescheduled to window end time. Let the refined deadline be represented as $timePt$. The time at which the pending tuples will be available is denoted as $ipAvailTime$. The time duration available for processing, $timeDur,$ is determined as the difference between $timePt$ and $ipAvailTime$. If it is possible to process all the pending tuples in the available duration, then processing can be done in a single batch as in case 3.  If the sufficient duration is not there, as in case 4, then the number of tuples that can be processed during this duration, $numTupleProc$, is determined. The remaining tuples must be processed before this point and hence the $timePt$ is again shifted. This process is repeated until all tuples are processed. Thus the function \textit{ScheduleWithoutAggCost} determines the batches and the size of each batch, $schTuples.$  Though a simple linear model without overhead cost is considered for explaining the concept, the scheduling schemes proposed work for any arbitrary cost model.

%In the static scenario, we have assumed that the input rate can be predetermined and the total number of tuples is known. But in many cases, the input rate and the total number of tuples may not be fixed. Handling of variable rate/number of tuples is discussed later in Section \ref{sec_variableIpRate}.

The batch size determined by the function \textit{ScheduleWithoutAggCost} incurs the least computation cost. The intuition behind the proof for the same is explained here. Let Algo-O represent the function \textit{ScheduleWithoutAggCost} proposed in this paper and Algo-C represent any arbitrary competitive algorithm. Let Batches of Algo-O vary from 1 to n1 and Batches of Algo-C vary from 1 to n2. For Algo-C to be better than Algo-O, n1 has to be greater than n2. For this, Algo-C has to process more tuples than Algo-O over a given duration. The last batch of tuples can be processed only at Window End or later. While Algo-O starts at Window End Algo-C may start at Window End or later. Thereby the duration of the last batch of Algo-O will be greater than or equal to that of Algo-C. Algo O tries to maximize the number of tuples processed in every batch. Thus the total number of tuples processed by Algo-C in the last batch cannot be greater than that processed by Algo-O. Therefore, the number of pending tuples to be processed before the last batch for Algo-O will be less than or equal to that of Algo-C. Input availability for Algo-O will be earlier or the same as that of Algo-C. Thus, for every batch Algo-C cannot be started earlier than Algo-O. Thus the total number of tuples processed by Algo-O over a given duration is always greater than or equal to that processed by Algo-C. Therefore the total number of batches, n1 will always be less than or equal to n2.

An alternative scheme for determining the optimal batch sizes based on linear arithmetic constraints, which is applicable only for the linear cost model is explained in Section \ref{single-query-constraints}.  We carried out several experiments using the two schemes, and found that the results obtained from both methods were the same in all the cases we considered. 

In the static scenario, we have assumed that the input rate can be predetermined and the total number of tuples is known. But in many cases, input rate may vary and the total number of tuples may not be fixed. We refer to such scenarios as dynamic scenarios and the scheduling scheme for the same is presented in the Section \ref{sec_variableIpRate}.

\subsection{Constraints based Scheduling using Optimizer}  \label{single-query-constraints}

Here a constraints-based approach is presented for determining the optimal batch size for linear cost models. As batch size increases, the number of batches reduces, and thereby the overall computation cost also reduces. Thus the problem can be formulated using a set of constraints to find out the minimum number of batches and the optimal batch size of each of these, such that all tuples can be processed within the given deadline. 

Each $x_{i}$ denotes a batch and its coefficient denotes the batch size. Let each Batch x$_{i}$ have a start time denoted as schStartTime. Also, the computation duration or cost of the batch is represented as schDur. Constraint 1 is defined once for a query. Further for each batch, constraints 2 and 3 are defined.

The constraints to be satisfied for scheduling are defined below.

    \begin{itemize}
    \item \textbf{Constraint-1:} The sum of the number of tuples of all the batches must be equal to the total number of tuples. This is given in (\ref{eqn_constraint-1.1}).  
    \item \textbf{Constraint-2:} Each batch must be processed before the scheduling of the next batch commences. The sum of schStartTime and schDur of Batch $x_{i}$, has to be less than or equal to the schStartTime of Batch $x_{i+1}.$ And for the last Batch $x_{n},$ this sum has to be less than or equal to the window deadline. These are given in (\ref{eqn_constraint-1.2}) and (\ref{eqn_constraint-1.3}).
    \item \textbf{Constraint-3:} The number of tuples available at the time point when the computation of a batch starts has to be greater than or equal to the sum of all the batch sizes including the i$^{th}$ batch. This is given in (\ref{eqn_constraint-1.4}). 
\end{itemize}
\begin{comment}
\begin{equation} \label{eqn_constraint-1.1}
\begin{split}
    numTupleTotal  = \sum_{i=1}^{n}x_{i} & \\
   schStartTime_{x_{i}} + schDur_{x_{i}} & <= schStartTime_{x_{i+1}} \\ where & \; i \neq n \\
   schStartTime_{x_{n}} + schDur_{x_{n}} <= & deadline_{window} \\
   inputRate * schStartTime_{x_{i}}  >= &\sum_{j=1}^{i}x_{j} \; \: where \; i = 1..n 
\end{split}
\end{equation}
\end{comment}

\begin{equation} \label{eqn_constraint-1.1}
\begin{split}
    numTupleTotal & = \sum_{i=1}^{n}x_{i} 
\end{split}
\end{equation}

\begin{equation} \label{eqn_constraint-1.2}
\begin{split}
   schStartTime_{x_{i}} + schDur_{x_{i}} & <= schStartTime_{x_{i+1}} \\ & where \; i \neq n 
\end{split}
\end{equation}

\begin{equation} \label{eqn_constraint-1.3}
\begin{split}
   schStartTime_{x_{n}} + schDur_{x_{n}} <= deadline_{window} 
\end{split}
\end{equation}

\begin{equation} \label{eqn_constraint-1.4}
\begin{split}
  inputRate * schStartTime_{x_{i}}  >= \sum_{j=1}^{i}x_{j} \; \: where \; i = 1..n 
\end{split}
\end{equation}

Using the interfaces provided in Google's OR-Tools \cite{ref_web_googleOr}, for Mixed-Integer Programming solvers, the above constraints were defined and solved. The number of batches is defined as an input to the optimiser module. The optimiser solves the given constraints for the defined number of batches. The solution with the least computation cost is the one that processes all the tuples using the least number of batches.

Consider case-3 and case-4 sample queries presented in section \ref{sec:singleq}. For these queries the number of batches, represented as i can vary from 1 to 10. The optimiser solved the case-3 query using 2 batches of size 6 tuples and 4 tuples respectively. Similarly, case-4 is solved in 3 batches of sizes 4, 4, and 2 tuples respectively. 
This is the same as the results obtained using the scheduling schemes proposed in Algorithm \ref{algo_sch_single_slots}.  

However, note that the constraints based model supports only the linear cost model,
but Algorithm \ref{algo_sch_single_slots} works on any arbitrary cost model.

\section{Dynamic Scenarios} 
\label{dynamic}

For the static single query case, as the requirements are known prior, it can be scheduled with the least possible computation cost as presented in Section \ref{static}. In reality, there may be multiple queries running on the system. These queries may have the same or different deadlines and may or may not use the same input data stream. A generic scenario where the system has to process multiple queries within their respective deadlines is referred to as a dynamic scenario in this paper and the scheduling scheme for the same is presented in this section.

There can be different uncertainties in the dynamic scenario. Queries may be added or removed from the system randomly and in addition, the input rates and the total number of tuples of queries may also vary. Handling these uncertainties is explained in this section.  Here, each of the queries is processed in multiple batches and the methodology of determining the batch size is explained in Section $\ref{dyn_batch_size_det}.$ 
Section $\ref{sec_dynamic_sch_scheme}$ explains the scheduling scheme for a system where new queries can be added or removed at any point, with the assumption that the input rates of the queries can be predetermined. The schedulability of queries under such scenario is given in the section $\ref{sec_schedulability}.$ Section $\ref{sec_variableIpRate}$ explains the scheduling scheme for the scenario where the input rates and the total number of tuples of queries vary. 

\subsection{Determining Minimum Batch Size} \label{dyn_batch_size_det}

Resource availability is ensured in the static single query scenario and hence, it is possible to determine the appropriate batch sizes which will incur the minimum computation cost. On the contrary, for the dynamic scenario, the  scheduler does not have apriori knowledge of queries. If query execution is delayed for the appropriate batch size, resource availability cannot be ensured as other queries can be added or removed from the system, affecting resource availability.  To handle the dynamic scenario, our approach is to process queries intermittently, whenever the number of tuples available for processing exceeds some minimum batch size. Aggregation on very small batches may not lead to benefits, for example, if each batch has only 1 tuple on average. On the other hand, it is better if the minimum batch size is greater than or at least twice the number of groups in the group by query, as there will be a  significant reduction in the number of intermediate tuples. Queries without Group By can be treated as a special case where the number of groups is one. Processing queries intermittently in batches ensures a reduction in overall computation cost compared to the approaches used in the existing stream processing engines, such as tuple-by-tuple or micro-batch based processing. 

Determining the minimum batch size for the dynamic case is a trade-off between minimising the overall computation cost and ensuring that query deadlines are met. While smaller batches incur more computation cost, larger batches may affect the deadline of other queries running on the system. 

Minimum batch size referred to as MinBatchSize, is determined based on the Resource Slack Factor $\delta_{RSF}$.  The goal is to pick a minimum batch size such that the overall computation cost does not exceed more than $\delta_{RSF}$ fraction compared
to the minimum computation cost of a single batch, i.e. $minCompCost_{BatchSize=x}$ $\leq \delta_{RSF}$  $*  minCompCost_{BatchSize=N}.$ N denotes the total number of tuples and $minCompCost_{BatchSize=N}$
denotes the minimum computation cost required for processing all
tuples in a single batch. $minCompCost_{BatchSize=x}$ denotes the minimum computation cost for processing all the tuples in multiple batches, where each batch is of size $x$ tuples. The lower bound of $minCompCost_{BatchSize=x}$ is $minCompCost_{BatchSize=N}.$ 
$\delta_{RSF}$ can be set based on the system utilisation. If the system is lightly loaded then a larger $\delta_{RSF}$ can be set. Extensions to  automatically adjust it based on the system load are part of future work.

\subsection {Scheduling using Minimum Batch
Size} \label{sec_dynamic_sch_scheme}

Once the MinbatchSize is determined, queries can be processed using any of the approaches such as Least Laxity First (LLF), Earliest Deadline First (EDF), Shortest Job First (SJF), and Round Robin (RR).

Unlike the traditional scheduling algorithms, for stream processing applications, while scheduling the queries, the availability of inputs has to be taken into account. For simplicity, Non-preemptive scheduling strategy is followed here, where each query runs to process its Minbatch tuples without any preemption.

If the system is idle, a new query request is honored immediately and its MinBatch is computed. If any batch is already under process when a new query with the least slack or earliest deadline is submitted, then the scheduler takes the new query at the end of the batch execution. Once a batch has been processed both CPU and memory is released. The intermediate results of the batch are stored in the form of files. This is unlike the streaming approach where the intermediate results are stored in memory. Storing intermediate results in memory can lead to out-of-memory issues, especially for long-running queries with joins. The same was observed with some of the TPC-H queries when run in streaming mode. The results of the experiments are presented in  Section~\ref{results}.

LLF based scheduling is explained here. For each new query added to the system, its Minimum Batch
size is computed. If the currently available tuples,
NumTupleCurr is greater than or equal to MinBatchSize, then laxity
is computed with respect to the current time, as per (\ref{eqn_laxity}).

\begin{equation} \label{eqn_laxity}
\begin{split}
    laxity_{Q_{i}}  =  deadline_{Q_{i}} - currentTime  
             - CompCost_{Q_{i_{(batchSize=x)}}} 
\end{split}
\end{equation}
The query with
the least laxity is given the highest priority and its Min batch is processed. 

\floatname{algorithm}{Algorithm}

\begin{algorithm}[t]
\caption{ScheduleDynamicLLF()} \label{algo_var_min_batch_size}
\begin{flushleft}
\begin{algorithmic}
\STATE \textbf{}
\STATE \textbf{Input:} 
\STATE \hskip0.3em $\delta_{RSF}$ $//Resource$ $slack$ $factor$
\STATE \hskip0.3em $C_{max}$ $//Max$ $allowed$ $computation$ $time$  $for$ $each$ $minbatch$
\STATE $qList[]$  $//list$ $of$ $queries$
\STATE $LARGE\_NUMBER$ $//sufficiently$ $large$ $number$ $greater$ $than$ $slack$ $time$
\STATE \textbf{repeat\{}
\IF{New Query}
\STATE $Add$ $New$ $Query$ $to$ $qList[qList.length]$
\STATE $qList[qList.length-1].MinBatchSize$ $=$ \\ \hskip1em $\textbf{FindMinBatchSize}($ $qList[qList.length-1],$ $\delta_{RSF}$, \\ \hskip10em $C_{max}$ $\textbf{)}$
\ENDIF
\IF{Delete Query}
\STATE $Remove$ $query$ $from$ $qList$
\ENDIF
\IF{$(qList.length>0)$}
\FOR{$(i=0;i<qList.length;i++)$}
\STATE $qList[i].SlackTime_{MinBatchSize}$ $=$ $LARGE\_NUMBER$
\STATE $qList[i].NumTupleCurr$ $=$ $\textbf{FindCurrTuples}(qList[i])$  
\IF{$qList[i].NumTupleCurr >=  qList[i].MinBatchSize$}
    \STATE $qList[i].MinCompCost$ $=$ \\ \hskip1.5em $\textbf{FindMinCompCost}(qList[i])$
    \STATE $qList[i].SlackTime_{MinBatchSize}$ $=$ \\ \hskip1.5em $\textbf{FindSlackTime}(qList[i],curTime)$
    \STATE $qList[i].CurBatchSize =  qList[i].MinBatchSize$
\ENDIF
\ENDFOR
\STATE $sort$ $qList$ $based$ $on$ $slack$ $time$
\STATE $process$ $qList[0]$ 
\IF{$(qList[0]$ $is$ $completed)$}
\STATE $Remove$ $qList[0]$
\ENDIF
\ENDIF
\STATE \textbf{\}for ever}
\STATE \textbf{END}
\end{algorithmic}
\end{flushleft}
\end{algorithm}

The scheduling logic is given in Algorithm \ref{algo_var_min_batch_size} \textit{ScheduleDynamicLLF}.  As and when new
queries are added to the system function \textit{FindMinBatchSize} in
Algorithm \ref{algo_var_min_batch_size} determines the MinbatchSize such that the overall computation cost including the final aggregation does not increase more than $\delta_{RSF}$ times the $minCompCost_{batchSize=N}$. While determining the Minimum Batch size for queries, it has to be ensured that, any batch of a query has to be processed within a time period denoted as $C_{max}$. Thus MinBatchSize is determined such that its minimum computation cost does not exceed $C_{max}$.  Since the scheduler is non-preemptive, $C_{max}$ ensures that any new query inputted to the system with the least slack is not delayed for processing by more than $C_{max}$. The value of $C_{max}$  has to be decided based on the application latency requirements. 

Then, for all the queries in qList, MinCompCost, and SlackTime or laxity is determined. Function \textit{FindMinCompCost} determines the minimum computation cost required for processing the pending tuples in multiple batches, where each batch is of MinBatchSize. Function \textit{FindCurrTuples} determines the number of tuples that are currently available for processing. Jitter in input rate can be handled as the processing of Minbatch is triggered based on the number of tuples available. Then, laxity is computed by the Function \textit{FindSlackTime} for those queries whose Minbatches are ready for processing. Then, the query with the least laxity is picked and processed. If all tuples of a query have been processed, then the final aggregation is done and the query is removed from $q_{list}.$
   
Similarly, EDF, SJF, and RR based scheduling can be done. Experiments were carried out with all these approaches and the results are given in Section 7. 

\subsection{Discussions on Schedulability} \label{sec_schedulability}

Determining the schedulability of a given set of tasks with Non Preemptive scheduling is known to be NP complete in general (see e.g. Georges et al.\ \cite{ref_np_edf}).  This is true even if we include the non-idling constraint, where an earlier deadline task that has been released, cannot wait unless the processor is currently processing another task.
A Non Idling Non Preemptive (NINP) EDF scheduler processes the earliest deadline task amongst all those that are released whenever the processor becomes free.  

While scheduling using Non Preemptive scheduling there may be scenarios where a task with an earlier deadline may wait as the scheduler is already occupied with another task with a later deadline that happened to be released earlier.  The duration required for completing the currently running task, referred to as the blocking period plays an important role in determining the feasibility of the schedule. Georges et al.\ \cite{ref_np_edf} show that with certain constraints on the blocking period,  where all tasks have the same duration,  NINP EDF is optimal for any sequence of n concrete tasks.  NINP-EDF is also known to be optimal under the periodic task case with certain constraints, but tasks are not periodic in our scenario.  

The dynamic scenario we have considered uses NINP EDF. In our scenario, each query is processed in one or more batches. A batch can be correlated to the task defined in Georges et al.\ \cite{ref_np_edf} model. Though we have not considered all batches to be of equal duration,  the execution duration is constrained such that it less than or equal to $C_{max}.$ Although NINP-EDF may not result in an optimal schedule in all the cases, it can be used as a heuristic.

%For a special case where the execution duration of all batches is equal, NINP EDF is optimal.

\subsection{Handling Variable Input Rate} \label{sec_variableIpRate}
So far we have assumed that the input data rate can be modelled and the total number of tuples is fixed. But in reality, both the input rate and the total number of tuples can vary or the total number of tuples may be fixed while the input rate varies. Handling these uncertainties is explained in this section. 

Consider the scenario where the total number of tuples is fixed, but the input rate varies. For such scenarios, during MinBatchSize determination, the expected time point at which MinBatchSize will mature as per the input rate is also estimated. Slack time is computed for queries whose input has reached MinBatchSize or the time point has crossed its estimated input available time. Then the queries are sorted based on their slack time and the query with the least slack time is scheduled for processing. If the actual input rate is faster than or equal to the predicted model then processing will be triggered as and when the required batch size is ready. If the actual input rate is slower, then processing gets triggered based the on the estimated input available time. In such cases, the proposed scheme tries to process the available tuples instead of waiting for the Minbatch readiness, thereby trying to meet the deadline. 

For dynamic systems where both the input rate and the total number of tuples can vary, an estimator can be used to derive the total number of tuples based on the actual input rate. Laxity is then computed for all queries irrespective of their MinBatch readiness, using the estimated total number of tuples. Then queries are sorted based on slack time and the query with the least slack time is processed. Thus the system is not idle and the available tuples are processed thereby reducing the risk of missing the deadline.

\section{Related Work}
\label{relwork}

Many stream processing engines run on the YARN infrastructure. Vavilapalli et al.\ \cite{ref_yarn} explain YARN and the scheduling schemes supported such as FIFO, Capacity, and Fair. YARN caters to generic resource management and job scheduling. YARN does not support scheduling queries with deadline constraints. Stream Processing Engines such as Apache Spark and Flink process tuples eagerly with some fixed minimum batch size. However, they do not consider deadlines, and do not utilise the available slack time which can allow for larger batches and thus lower cost. Ye et al. \cite{ref_nostop} propose an optimization technique for Spark Streaming configurations without considering deadlines.

Shang et al.\ \cite{ref_crocodile_pap}, Tang et al.\ \cite{ref_iqp},  address cost reduction in scenarios not bound by absolute deadlines. On the other hand, Ou et al.\ \cite{ref_tick}, Li et al.\ \cite{ref_edf_streamproc1}, focus on only meeting deadline constraints and do not talk about cost optimisation. The Scheduling schemes proposed by us in this paper primarily focuses on scheduling queries within deadline and in addition tries to minimise the overall computation cost by determining the appropriate batch size which will incur less computation cost while meeting the deadline constraints. 

Tang et al.\ \cite{ref_iqp}, Shang et al. \cite{ref_crocodile_pap} point out that the slack period available in queries can be utilised to reduce resource consumption. Present Stream Processing Engines consume resources continuously. Tang et al.\ \cite{ref_iqp} propose Intermittent Query Processing where tuples are processed intermittently.  While processing all tuples in a single batch would incur the least cost, it may not be feasible to meet the deadline. So part of the work that supports incremental operations is done intermittently. The intermittent query processing is triggered at some time interval or based on the number of tuples accumulated. Shang et al.\ \cite{ref_crocodile_pap} have built a database system namely CrocodileDB which processes queries intermittently based on user inputted frequency. But these queries are not bound to be completed within an absolute deadline. Not only are our queries deadline bound, but also we propose a scheduling strategy that meets the deadline while processing the queries with minimal computation cost.  

Tang et al.\ \cite{ref_thriftyqp} define a new metric, Incrementability, to denote the amount by which a query supports incremental operations. The given query is broken down into query paths or subqueries where incrementable subqueries are processed more frequently compared to others.  Each path gets processed when an estimated set of tuples arrive and this is denoted as a pace configuration. Pace configuration is estimated using a greedy algorithm, where the total work done in processing the query is minimised while ensuring that the work done after the last tuple arrives is not beyond the user specified constraint. 

Here, the absolute deadline of the query is not inputted into the system. Instead, the user inputs a final work constraint which denotes the amount of work that can be done after the last tuple arrives. Though this is comparable to the static scenario considered in our paper, the major difference in the problem statement between ours and Tang et al.\ \cite{ref_thriftyqp} is that while our queries have to be completed within some absolute time, in the case of Tang et al.\ \cite{ref_thriftyqp}, there is no absolute time within which the final work has to be completed. 

Also, our algorithm automatically determines the amount of work that can be done after the last tuple arrives and also schedules the pending work prior to the window end by determining the appropriate batch sizes. Also, unlike the pace configuration where each query path has a fixed batch size, batch size in our algorithms is varied and maximized to achieve minimum computation cost. In addition, while Tang et al.\ \cite{ref_thriftyqp} do not address any scheduling strategies for multi query scenarios, we have addressed a dynamic multi query scenario in our work. We have proposed an EDF/LLF based scheduling strategy for processing multiple queries in Minbatches to achieve cost reduction while honouring the deadline constraints. Experimental results have shown that our approach was able to meet the deadlines whenever a feasible schedule existed. 

Tang et al.\ \cite{ref_resourceEffiQryExe} talk about a shared query plan in the multi query scenario with different scheduling frequencies, which is an augmentation of their above work.  Cost reduction using multi query optimisation is planned as part of our future work. 

Wang et al.\ \cite{ref_grosbeak} have designed Grosbeak to schedule routine analysis jobs in non peak hours based on history of resource utilisation. The job is processed in batches which is similar to our approach, but details on scheduling are not discussed in \cite{ref_grosbeak}. Wang et al.\ \cite{ref_tempura} discuss optimisation of intermittent query processing, but unlike our case, they do not consider absolute deadlines or scheduling. 

Zhang et al \cite{ref_intraQryJoin} experimentally show that join queries processed in lazy manner can perform better than eager processing, but deadlines are not considered here. 

Some data stream systems, model each tuple with a deadline. Such models are required in hard real time systems where the incoming tuple has to be processed within a certain time to take  critical decisions. This is different from our problem statement where all tuples in a query have the same deadline and their results are required only at a certain point in time later. Ou et al.\ \cite{ref_tick} proposes Tick scheduling for such models. Tick denotes a set of tuples that have the same deadline. Tuples belonging to the same ticks are processed together based on the Earliest Deadline First (EDF) strategy. Li et al.\ \cite{ref_edf_streamproc1}, \cite{ref_edf_streamproc2} also talk about scheduling using EDF strategy. Unlike our approach, Ou et al.\ \cite{ref_tick}, Li et al.\ \cite{ref_edf_streamproc1}, \cite{ref_edf_streamproc2} do not talk about minimising computation cost by utilising slackness. Also, these models do not guarantee to process all tuples within their deadlines and overdue tuples are dropped.  

Scheduling in real time systems has been widely explored \cite{ref_hardRTSch_book1} and \cite{ref_realTimeSys_book2} and some of the prominent algorithms available in the literature are EDF, LLF, etc. While EDF and LLF scheduling only aim at completing the query within its deadline, our approach in single query scheduling determines the appropriate batch sizes which incur the least computational cost. In the multi query scenario, we aim to reduce the overall computational cost of each query by processing queries in Minbatches.

Existing stream processing engines do not account for deadlines, but instead focus on latency. Tuning of batch size is done to achieve a balance between throughput and latency, but not in a deadline aware manner, unlike our work. On the other hand, deadline aware scheduling algorithms (see for example the survey \cite{ref_schComp}) do not consider batching. To the best of our knowledge, our work is the first of its kind which combines batching and scheduling to honour deadline and minimise cost.

Apart from the above, there has been work in reducing the resource consumption in a cluster environment under deadline bound scenarios.  
Dimopoulos et al.\ \cite{ref_justice} allocate optimal resources based on past execution logs. Sidhanta et al.\ \cite{ref_optex} and  Islam et al.\ \cite{ref_dspark} define an objective function with factors such as cost and number of executors. This function is then minimised while meeting the deadline constraint. Cheng et al.\ \cite{ref_adaptiveSchSpark}  categorizes jobs as independent and dependent ones, where the independent jobs are scheduled using Fair scheduler while dependent jobs are scheduled using FIFO scheduler.

Wang et al.\ \cite{ref_hard_real_time_yarn} describe a scheduling approach in the YARN context, where in addition to a deadline and execution time, tasks have a value and a time-varying value density which are used for scheduling.  They do not consider batching to optimize resource consumption.

%These works focus on optimising the resource allocation, since adding more processing nodes may not be beneficial beyond some point.
In all these approaches, once the resources are allocated, either Tuple-by-tuple processing or micro batch processing happens based on the underlying Stream Processing Engine used.  They do not consider the issue of batch size or scheduling of batches, unlike our work. In our scheduling approaches, cost reduction is achieved by utilising query slackness. %Designing strategies for resource allocation in a cluster environment where more than one query can be scheduled in a cluster simultaneously is part of our future work. 

\section{Implementation of Custom Query Scheduler over Spark} 
\label{sec_implementation}

Apache Spark is an open-source distributed platform for carrying out data processing. It supports both batch processing and stream processing. Stream processing is achieved by processing real-time data in micro batches. The scheduling schemes proposed in this paper have been implemented by building a layer over the Spark architecture. The implementation aspects are described in this section.

\subsection{Overview of Custom Query Scheduler}\label{sec_cqs}

Spark architecture comprises a Driver program, Cluster Manager, and Worker nodes. We implement a Custom Query Scheduler on top of Spark. A typical Spark Application contains a Driver program that creates a Spark Context and performs the required operations by coordinating with the Cluster manager. Query scheduling is carried out by the Cluster manager which can be Spark's standalone cluster or a resource manager such as YARN, MESOS, etc.

In this model, the proposed scheduling algorithms are implemented in Custom Query Scheduler (CQS) which resides inside the Driver program. Let the Query details be fed to this CQS. In our implementation, records are stored in files that get consumed by the system at a predefined rate.  File based inputs are widely used in many systems. However, we have also done experiments using Kafka based input and the same is presented in section \ref{sec_kafka_based_proc}. 

Our scheduler can run in two modes - Single Query Mode and Multi Query Mode. CQS consists of a Query Repository, Schedule Optimizer and Query Scheduler components. The Query Repository consists of the queries along with its metadata. Also, as the current implementation uses Spark, it consists of the spark operations which are executed for each batch and the ones which are executed as part of the final aggregation. The Schedule Optimizer determines the appropriate batch sizes. Also it keeps track of the batches processed and invokes the final aggregation once all batches of a query have been processed. The Query Scheduler checks for any new input and schedules its processing with the help of the Schedule Optimiser.

In the Single Query Mode, one query at a time is fed to CQS. CQS computes the optimal batch sizes.  We have used TPC-H Dataset and the details of the same is explained in Section \ref{sec_tpc_dataset}. Each file of Orders is about 1.2 MB and each file of Lineitem is about 5 MB. Other relations such as Customer, Parts, Parts Supplier, etc are considered to be static information that do
not change during query execution. CQS is aware of the input file name and its location. Files for a batch are put into a sub-directory and the appropriate function which carries out the Spark Operations is invoked, to evaluate the given query. 

The intermediate results of each batch are stored in a file. Once all batches have been processed, another function is invoked to perform the final aggregation from these intermediate results. In the case of queries where the average has to be computed, as part of intermediate results sum and count are stored. During the final aggregation, using this sum and count, the average is computed. %Though currently, Spark operations have to be written in such a manner to take care of the final aggregation, input queries can be analysed and rewritten in an automated manner, to compute the intermediate and final results. 

In the Multi Query Mode, the scheduler in the Custom Optimiser runs periodically, at an interval of $C_{max}$, performing a set of predefined tasks as explained in Section \ref{dynamic}. CQS schedules the queries in time sharing mode using any of the scheduling strategies such as EDF, LLF, SJF, or RR.  

CQS checks whether the Minbatch of each query is ready for processing. Similar to the single query scheduling approach, CQS knows the input file name and its location. Separate sub-directories are created for each Minbatch. The query to be processed is selected based on the scheduling strategy  and appropriate functions are invoked to perform the Spark Operations.  Once all the tuples of a query have been processed, final aggregation is done in the last batch. To reduce the overhead of Spark Context creation for every Spark job, a common Spark Context is created by the Scheduler and is shared across all Spark jobs.

For handling queries with joins the following strategy is adopted. There can be two types of joins: stream-to-stream and stream-to-static data. For the stream to static join, in the proposed method, it is assumed that the static data does not change over the period of query execution. Hence each batch is joined against the  static data to get the join results. For stream-to-stream joins, joins can be performed across pairs of batches that would contain tuples that may
potentially be joinable. For simplicity, our implementation assumes that the corresponding tuples will be available in the same batch. For example, consider the join between orders and lineitem from TPC-H Dataset. It is reasonable to expect the order and its associated lineitem tuples to arrive in the same batch.

Apache Spark provides Spark Streaming and Structured Streaming \cite{ref_web_spark} for processing streaming inputs. In Spark Streaming, input tuples are processed in micro batches. In Structured Streaming, input data is considered to be part of an infinite table. 
%While Spark Streaming operates on RDDs, Structured Streaming uses Dataframes and Datasets. 
Structured streaming also uses a micro batch processing engine and operates at the trigger interval. Input tuples get added at the rate of trigger interval and get processed thereby updating the results. Results when outputted in the Complete mode give the final aggregated values.

\begin{figure}[tb]
\centering
\fcolorbox{black}{white}{
\def\svgwidth{0.45\textwidth}
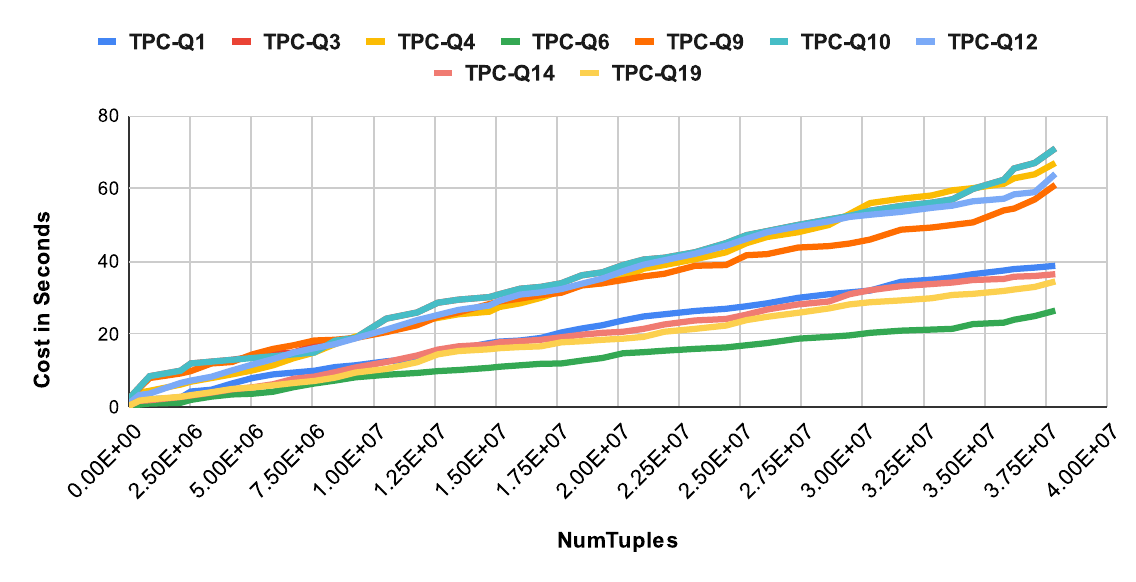}
\caption{TPC-H Queries Cost Model} 
\Description{TPC-H Queries Cost Model} 
\label{fig:tpc-cost-model}
\end{figure}

\subsection{Cost Modelling} \label{sec_cost_modelling}

 Each query is executed individually to experimentally measure the execution time versus input size. 
  
 The input consists of a total of 4500 files of Orders and Lineitem comprising 25GB of data. Input files are stored in the file system,  and queries are processed by invoking the appropriate Spark operations on these files. 

The number of Order tuples processed is varied in steps of 2000K, and the number of corresponding Lineitem tuples is as per the TPC-H benchmark. Queries are run on 2 Intel Xeon Silver 4116 Processors (2.10GHz) with 250GB of RAM. Spark context was configured with 48 cores and 20 GB of memory. 

The execution times of the different queries are shown in Figure \ref{fig:tpc-cost-model}. The cost model here maps batch size to cost for a fixed input size. In a purely linear model, cost per tuple does not depend on batch size but there can be an overhead per batch, while in a sublinear case, cost per tuple reduces with batch size. From the experimental logs the cost variation with respect to the batch size was analysed and fitted into the piece-wise linear cost model.

As queries will be processed in batches for both static and dynamic scenario, the intermediate results of each batch are written to a file which gets processed at the end to get the final aggregated  results. 

We model the final aggregation cost as follows. 
The final aggregation cost depends on the number of tuples, the number of batches in which it is processed, the number of groups in each batch, and the total number of groups.  The number of groups in a batch is a sub-linear function of the input size, but we do not include it in our model for the final aggregation cost, which is based only on the number of batches.   We fit a piece-wise linear model to estimate the final aggregation cost.

\subsection{Kafka based Spark Processing} \label{sec_kafka_based_proc}

File based input is widely used for bulk data since it enables easy information exchange, but streaming data platforms such as the widely used Kafka platform, are an alternative.  In this section, we explore input from a Kafka system.  For processing in the Batch mode, we get the count of the messages available using GetOffsetShell from Kafka tools and then trigger processing if sufficient tuples are available by setting the appropriate starting and ending offsets. While reading from Kafka in the stream mode, if all tuples are available, the Trigger once option can be used which reads all the tuples from Kafka and does the processing in one go. 

To measure performance, two Kafka topics namely \textit{orders} and \textit{lineitem} were created with 36 partitions to support parallel processing.
We explored using Kafka in three ways: (i) in the normal streaming mode using readstream(), (ii) streaming using readstream(), but with trigger once option (which we refer to as OneShot), and (iii) batch mode using read() (which reads all tuples at one go, and which we refer to as Kafka Batch). Custom queries CQ1 to CQ4 were executed in all these modes, in addition to being executed using files as inputs instead of Kafka.  In the batch mode, we have processed all tuples in a single batch.

The results are shown in Table\ref{tbl:kafka_results}. It was observed that the cost incurred using Kafka is generally more than using file based inputs. Also Kakfa streaming incurs considerably more cost than using any of the batch/OneShot modes.  Thus, processing in batch mode significantly reduces the cost incurred compared to stream processing, whether we use files as input, or the Kafka platform.

\begin{table}[tb]
  \caption{Cost Comparison of Custom Queries}
  \label{tbl:kafka_results}
  \begin{tabular}{|l|llll|}
    \toprule
     \textbf{Description} & \textbf{CQ1} & \textbf{CQ2} & \textbf{CQ3} & \textbf{CQ4}\\
    \midrule    
    Kafka Streaming Default & 90.22 & 914.0 & 595.0 & 1428.0 \\ \hline
    KafKa Streaming OneShot & 63.0	& 67.6	& 246.9	& 379.4 \\ \hline
    Kafka Batch (NumBatch=1) & 57.0 & 57.8 & 191.33	& 266.0  \\ \hline
    File Based (NumBatch=1) & 17.86 &	18.85	& 32.0 &	32.5 \\ 
    \bottomrule
  \end{tabular}
\end{table}

\section{Performance Evaluation} \label{results}
In this section, we first briefly describe the dataset and the queries used. We then present a performance evaluation on different queries, focusing on the variation of computation cost w.r.t number of batches and then evaluate the Single and Multi Query mode scheduling strategies.       %A comparison of scheduling using other strategies apart from LLF is also presented in this section.

\subsection{Data and Queries} \label{sec_tpc_dataset}

We use a slightly modified version of the TPC-H Dataset, which  models a business environment.  Data is organised in eight tables, occupying a  total of 25GB.  In order to simulate the input data stream, a timestamp has been added to each record in Orders and Lineitem, which are considered as streaming inputs. 

As explained in section \ref{sec_implementation} our scheduling algorithms can be used for both file based and Kafka based inputs. For our performance studies to find the cost reduction using scheduling strategies, we used file based inputs to avoid overheads encountered in streaming data from Kafka.

The input stream consists of 4500 files inputted at the rate of 1 file of Orders and one file of Lineitem per second. Each file of Orders is about 1.2 MB and each file of Lineitem is about 5 MB. Other relations such as Customer, Parts, Parts Supplier, etc are considered to be static information that does not change during query execution.

A subset of TPC-H queries, including queries with joins has been considered. For example, here TPC-H Q4 performs a join on Orders and Lineitem which models a stream versus stream join.   While in general, a stream-to-stream join may need to match tuples that are arbitrarily separated in timestamp order, real-world stream-to-stream joins are usually correlated in time.  As mentioned earlier in Section~\ref{sec_cqs}, we assume that matching tuples between Orders and Lineitem are within the same input batch.

\begin{table}[t]
\caption{Queries for Evaluating Single \& Dynamic Scenario}
\begin{center}
\begin{tabular}{|l|l|}
\hline
\textbf{QueryID}&\textbf{Query} \\
\hline
    CQ1 & \vtop{\hbox{\strut \textbf{$SELECT$ $count(*)$ $as$ $totalOrders$ }}  \hbox{\strut \textbf{$FROM$ $orders$}}} \\
    CQ2 & \vtop{\hbox{\strut \textbf{$SELECT$ $count(*)$ $as$ $totalOrders,$ $orderPriority$}}  \hbox{\strut \textbf{$FROM$ $orders$ $GROUP$ $BY$  $orderPriority$}}} \\
    CQ3 & \vtop{\hbox{\strut \textbf{$SELECT$ $count(*)$ $as$ $totalItems,$ $suppKey$}}  \hbox{\strut \textbf{$FROM$ $lineItems$ $GROUP$ $BY$ $suppKey$}}} \\
    CQ4 & \vtop{\hbox{\strut \textbf{$SELECT$ $count(*)$ $as$ $totalItems,$ $partKey$}}  \hbox{\strut \textbf{$FROM$ $lineItems$ $GROUP$ $BY$ $partKey$}}} \\
\hline
%\multicolumn{4}{l}{$^{\mathrm{a}}$Sample of a Table footnote.}
\end{tabular}
\label{tbl:dyn-query-attr-exp}
\end{center}
\end{table}

On the other hand, TPC-H Q19 performs a join of Lineitem against Parts which represents the join of a streaming input with static data. TPC-H Q10 performs the join of two streams, as well as static data i.e. join is performed on Orders, Lineitem, Customer, and Nation. %TPC-H query, Q4 is shown here 

Apart from these TPC-H queries, some custom queries, as in Table \ref{tbl:dyn-query-attr-exp}  were also designed.  These are sample queries on incoming Orders and Lineitem streams for doing simple analytics. These queries do not have any filtering and do aggregation on the entire dataset of 25GB.  While all the custom queries compute count over the entire dataset, CQ1 does not have Group By and on the other hand, CQ2, CQ3, and CQ4 compute count with Group By. The number of groups for CQ2, CQ3, and CQ4 is around 5, 360K, 1500K respectively.

\subsection{Cost variation w.r.t. Number of Batches}

\begin{figure}[tb]
\centering
\fcolorbox{black}{white}{
\def\svgwidth{0.48\textwidth}
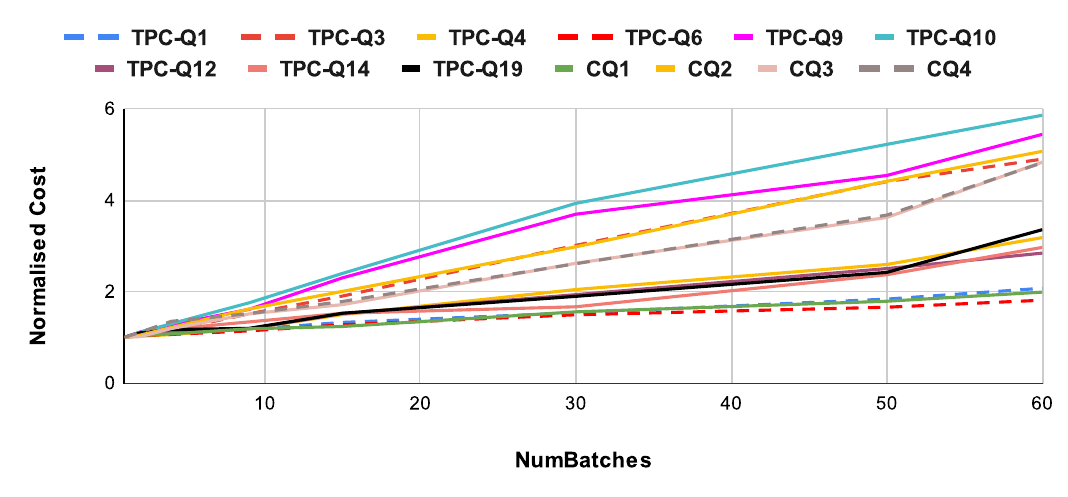}
\caption{Cost versus Number of Batches} 
\Description{Cost versus Number of Batches} 
\label{fig:costVsbatchTPC-Cust}
\end{figure}

Processing all tuples/files in a single batch incur the least computation cost. In our scheduling schemes, a batch is started once all the input tuples are available, so we measure the time to complete the processing of each batch. Cost then refers to the sum of the query execution time of all batches. The effect of the increase in computation cost as the number of batches increases is analysed here. For a Batch size of x, Spark Operations are invoked when x number of files are ready for processing. This is repeated until the set of 4500 files is processed. For example, for a batch size of 75, for every 75 files, Spark Operations are invoked. It takes 60 batches to process 4500 files and an additional aggregation task to get the final result. Similarly, queries were evaluated for different batch sizes of 2250, 1125, 500, 300, 150 and 90.

For each of the queries, the minimum cost required for processing it in a single batch is taken as the baseline. The cost incurred when
processed with different batch sizes has been normalised w.r.t this baseline as in Figure \ref{fig:costVsbatchTPC-Cust}. It can be observed that the more the number of batches, the more the overall computation cost. For example, consider TPC-H Q10, the computation cost  incurred when processed in 60 batches is almost 6 times the baseline cost. 

Also, the final aggregation cost increases with the increase in the number of groups in Group By, the number of batches, and tuples. This can be well observed in the Custom queries. For example with 60 batches, the computation cost of CQ2 is 2.7 times more than that of CQ1. As the number of batches and tuples are the same for CQ1 and CQ2, the difference in computation cost can be attributed to the more number of groups in CQ2 compared to that of CQ1. CQ3 and CQ4 operate on lineitem and have approximately 360K and 1500K number of groups respectively. When processed in 60 batches, the aggregation cost of CQ3 and CQ4 was approximately 3 and 7 respectively. For the same number of batches and the same number of tuples, CQ4's aggregation cost is higher than that of CQ3 as the number of groups in CQ4 is higher than that of CQ3. But there is only a slight increase in the overall computation cost of CQ4 when compared to CQ3. This is because the computation cost for processing per tuple is higher in CQ3 than in CQ4. 

\begin{figure}[tb]
\centering
\fcolorbox{black}{white}{
\def\svgwidth{0.48\textwidth}
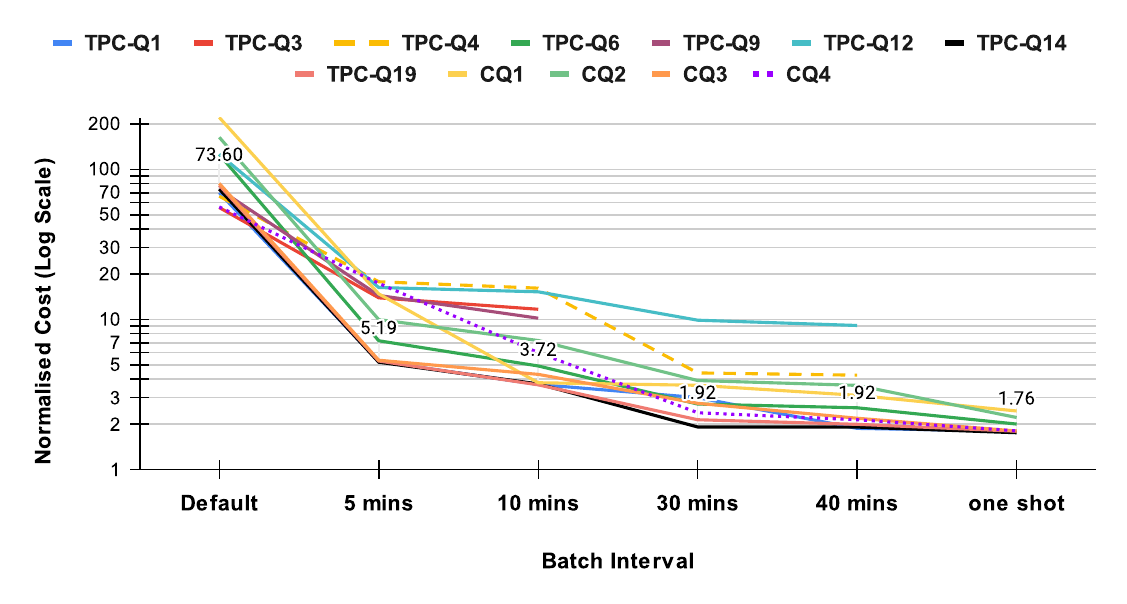}
%\textit{Bar bordered in Black outline executed with reduced window duration}
\caption{Comparison of Cost in Batch Mode versus Spark Streaming for Single Query Scenario} 
\Description{Comparison of Cost in Batch Mode versus Spark Streaming for Single Query Scenario} 
\label{fig:singleQuerySparkComp}
\end{figure}

To compare the efficiency of the processing in batches as against Spark Streaming, the queries were processed using the Streaming job in Apache Spark with different batch intervals of 5, 10, 30 and 40 minutes. In addition, experiments were done with the default batch interval, i.e. immediate and one-shot where all files are processed in one go. In case of default batch interval, Spark schedules each batch as soon as it completes the previous one. The results are compared against the cost incurred when processed using our method in a single batch. Results are shown in  Figure \ref{fig:singleQuerySparkComp}.

It can be observed from Figure \ref{fig:singleQuerySparkComp}, that the computation cost decreases as the batch interval increases. The least computation cost incurred by Spark streaming is with the one-shot mode of processing. Among all the queries, TPC-Q14 (data labels marked in the figure) has the least normalised computation cost and it is 1.76 times more than the cost incurred when all tuples are processed in a single batch using our approach. 

In Spark Streaming, watermarks and time constraints are defined for joins with stream data. This is because a stream represents an infinite input and hence for performing join some interval has to be defined. For joins in long-running queries, Spark Streaming may not be successful due to memory limitations. For experiments, Window duration was set as 4500 sec which corresponds to the total duration of the incoming stream.

For TPC-Q10, execution using Spark Streaming gave an out-of-memory error for the default batch interval. On increasing the Spark memory from 20GB to 45GB TPC-Q10 executed successfully.  But further on increasing the batch interval, the runs failed even with a 45GB memory configuration. On the other hand, using our approach, TPC-Q10 ran successfully considering the entire duration of 4500 sec with 20GB memory. Similarly, it was observed that TPC-Q3 and TPC-Q9 failed for a batch interval of 30, 40 minutes due to memory issues. Also, processing all tuples in one shot was not successful for TPC-Q4 and TPC-Q12. 

All these failed queries have join on Orders and Lineitem. Spark used Streaming Symmetric Hash Join for doing the join operation. Although streams are unbounded in size, watermarks and event time constraints allow joins to be done using only a part of each stream. In our experiments, watermarks and the event time constraint were both set to 2 sec.  It was observed that execution with an aggregation window interval of 30 and 40 minutes worked with the default batch interval, but the execution failed when the batch interval was increased to 30 and 40 minutes. As the batch interval increases in Spark Streaming, the amount of data to be processed in a batch increases, thereby demanding more memory. Zhang et al.\ \cite{ref_intraQryJoin} talk about similar findings where eager join algorithms are memory bound and thus require more memory. These cases failed even with the increased Spark memory of 45GB.  In contrast with our approach all queries executed successfully with 20GB memory.

To compare the computation cost incurred under multi query mode, the following experiment was carried out. For Spark Streaming, all queries could not be run simultaneously due to a memory issue. Hence multiple runs were simulated where as part of each run, a set of queries were executed. Run-1 consisted of queries CQ1, CQ2, CQ3, CQ4, TPC-Q4, and TPC-Q6. Run-2 comprised TPC-Q1, TPC-Q3, and TPC-Q14, and Run-3 comprised TPC-Q9, TPC-Q12, and TPC-Q19. TPC-Q10 with a reduced window duration of 2400 sec was part of Run-4, as 4500 window duration with spark memory of 20GB was failing due to out-of-memory error. 

The cost incurred in each of the runs was summed up to get the total computation cost. Spark streaming experiments were done using the default and 10 minutes batch intervals, as for larger intervals queries TPC-Q3, TPC-Q4, TPC-Q9, and TPC-Q12 failed. This is compared against the total computation cost incurred using our dynamic mode of scheduling with 50\% $\delta_{RSF}$ factor using LLF. The cost incurred by Spark streaming for default and 10 minutes batch intervals was 60 and 12 times respectively more compared to that of our approach. Also, Spark streaming does not guarantee to meet query deadlines. Thus our approach not only reduces the computation cost but schedules queries to meet their deadlines. Evaluation of our approach with respect to meeting deadlines is explained as part of sections \ref{res_single_query} and \ref{res_multi_query}.

%Spark streaming experiments were done using the default batch interval, as for larger intervals queries TPC-Q3, TPC-Q4, TPC-Q9, and TPC-Q12 failed. This is compared against the total computation cost incurred using our dynamic mode of scheduling with 50\% $\delta_{RSF}$ factor using LLF. The cost incurred by our approach was 28 times less compared to that of Spark streaming with default batch interval. But, Spark streaming does not guarantee to meet query deadlines. Thus our approach not only reduces the computation cost but schedules queries to meet their deadlines. Evaluation of our approach with respect to meeting deadlines is explained as part of sections \ref{res_single_query} and \ref{res_multi_query}.

%Spark streaming experiments were done using the default and 10 minutes batch intervals, as for larger intervals queries TPC-Q3, TPC-Q4, TPC-Q9, and TPC-Q12 failed. This is compared against the total computation cost incurred using our dynamic mode of scheduling with 50\% $\delta_{RSF}$ factor using LLF. The cost incurred by our approach was 28 and 6 times less compared to that of Spark streaming for default and 10 minutes batch intervals respectively. But, Spark streaming does not guarantee to meet query deadlines. Thus our approach not only reduces the computation cost but schedules queries to meet their deadlines. Evaluation of our approach with respect to meeting deadlines is explained as part of sections \ref{res_single_query} and \ref{res_multi_query}.

\subsection{Single Query Scenario} \label{res_single_query}

\begin{figure}[tb]
\centering
\fcolorbox{black}{white}{
\def\svgwidth{0.4\textwidth}
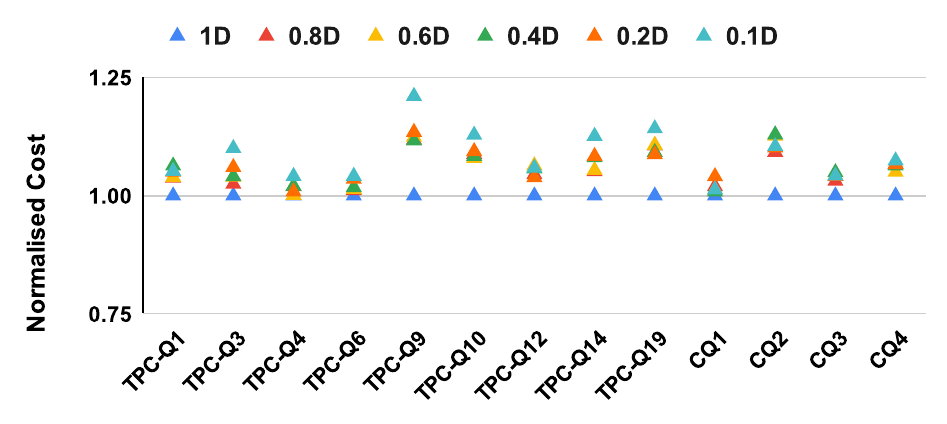}
\caption{Comparison of Cost in Single Query Scenario with different Deadlines} 
\Description{Comparison of Cost in Single Query Scenario with different Deadlines} 
\label{fig:singleQueryDeadline}
\end{figure}

TPC-H and Custom queries were evaluated with different deadlines. Here, the minimum computation cost required for processing the query in a single batch, after the Window End Time is taken as Deadline, 1 D. This Deadline was reduced to different levels as  0.8D, 0.6D, 0.4D, 0.2D and 0.1D time after the window end time.

The Query details are fed into the CQS. CQS determines the optimal batch sizes as per Algorithm \ref{algo_sch_single_slots}.
CQS then monitors the readiness of the batch for processing. %For wider deadlines, the size of the last batch increases, and thereby more files can be processed in the last batch. 
Once a batch is ready, the respective Spark Operations are invoked. If the number of batches exceeds one, then the final aggregation is also done. 

The result is shown in Figure \ref{fig:singleQueryDeadline}. The time taken in each case, normalized to the time taken using a single batch (1D) is shown in the figure. All cases were successfully executed within their deadlines. 

As the deadlines were reduced, it was observed that the number of tuples scheduled for processing after Window End decreases. This is as expected and it is due to the reduction in the time duration between the window end and the deadline.  For all deadlines except 0.1D, all queries were processed in two batches. For 0.1D, it was observed that queries TPC-Q3, TPC-Q9 and TPC-Q10 were processed in three batches while the other queries were processed in two batches. As shown in Figure \ref{fig:tpc-cost-model}, among all the queries, TPC-Q3, TPC-Q9, and TPC-Q10 queries have comparatively higher (especially in the  range of 0 to 500 in the x-axis) costs w.r.t other queries. Hence with the reduction in the deadline, for the 0.1D case, these queries were processed in more batches compared to other queries.

\begin{figure}[tb]
\centering
\fcolorbox{black}{white}{
\def\svgwidth{0.45\textwidth}
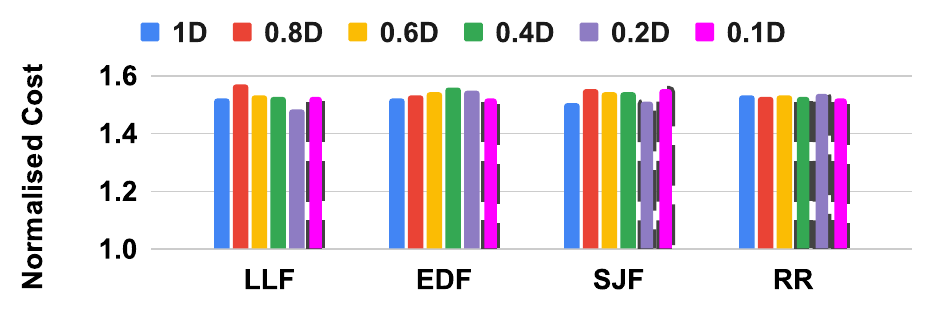}
\newline
\textit{\small Bars bordered in dashes denotes Deadline Missed cases}
\caption{Comparison of Cost in Multi Query Scenario with different Deadlines and scheduling Strategies}
\Description{Comparison of Cost in Multi Query Scenario with different Deadlines and scheduling Strategies}
\label{fig:multiQueryDeadlineFR}
\end{figure}
  
\begin{figure}[tb]
\centering
\fcolorbox{black}{white}{
\def\svgwidth{0.45\textwidth}
\input{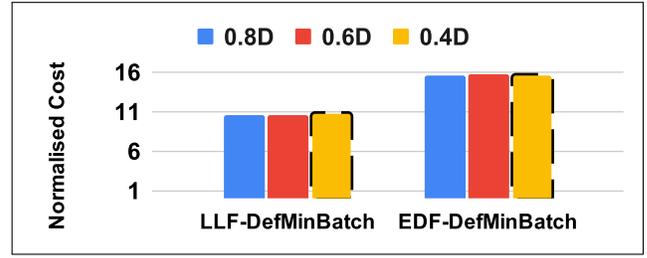}}
\newline
\textit{\small Bars bordered in dashes denotes Deadline Missed cases}
\caption{Comparison of Cost in Multi Query Scenario using scheduling Strategies without MinBatchSize}
\Description{Comparison of Cost in Multi Query Scenario using scheduling Strategies without MinBatchSize}
\label{fig:multiQueryDeadlineFR_Trad_EDFLLF}
\end{figure}

\subsection{Dynamic Scenario with Multiple Queries} \label{res_multi_query}
As in the single query case, for the dynamic scenario, query details are fed into CQS. Along with the query details, Resource Slack Factor, $\delta_{RSF}$ and $C_{max}$ are also submitted. CQS computes the Minbatch size for each query, based on the given cost model and  $\delta_{RSF}$ factor, such that the computation cost incurred in processing in batches does not exceed $\delta_{RSF}$ factor. Also, the computation cost for each batch must be less than or equal to $C_{max}.$ Then scheduling is done as explained in Section \ref{dynamic}.

Here all the TPC-H and the custom queries were considered simultaneously to depict the multi-query scenario.  $\delta_{RSF}$ factor of 50\% and $C_{max}$ of 30 seconds was considered. For each query, its Minbatch size was derived by CQS, based on the $\delta_{RSF}$ factor.  Amongst all queries, TPC-Q6 and CQ1 have the least Minbatch size, while queries CQ3 and CQ4 has the largest Minbatch size. The smallest and the largest Minbatch sizes are 180 and 1125 respectively. Thus TPC-Q6 and CQ1 gets processed in 25 batches while CQ3 and CQ4 get processed in 4 batches.

To evaluate the performance of the proposed method, Input arrival time and deadline are varied to generate different data sets as follows 
\begin{itemize}
    \item Choose a random number as Window Start Time
    \item Based on the input rate and the total number of tuples, the Window End Time is set accordingly.
    \item Let one of the queries be chosen randomly and its deadline shall be set as \[deadline_{Q_{1}} =  windEndTime_{Q_{1}}  + \] 
    \[ \hskip4em \delta * compCost_{Q_{1}{numbatch=1}}  + C_{max} \]
    \item Also it has to be ensured that there exists sufficient time to do the processing if the deadline overlaps with queries. For this, deadlines of the other queries are staggered as 
    \[IF(windEndTime_{Q_{i}}  >  deadline_{Q_{i-1}}) \]
             \[deadline_{Q_{i}}  =   windEndTime_{Q_{i}}  + \]
             \[\hskip4em \delta *  compCost_{Q_{i}{numbatch=1}} 
             + C_{max} \]
            \[ELSE \: deadline_{Q_{i}} =   deadline_{Q_{i-1}}  + \]
            \[ \hskip4em \delta * compCost_{Q_{i}{numbatch=1}} \: \forall i= 2 \:to \:n  \]
\end{itemize}

Let $\delta$ be a factor applied to the computation cost. By varying $\delta$, query slackness can be increased or reduced.

\begin{comment}

\begin{equation} \label{eqn_deadline_constraint1}
\begin{split}
    deadline_{Q_{1}} =  windEndTim&e_{Q_{1}}  +  \\ \delta * compCo&st_{Q_{1}{numbatch=1}}  + C_{max}  
    \end{split}
\end{equation}

\begin{equation} \label{eqn_deadline_constraint2}
    \begin{split}
        IF(wi&ndEndTime_{Q_{i}}  >  deadline_{Q_{i-1}}) \\
             dead&line_{Q_{i}}  =   windEndTime_{Q_{i}}  + \\    &\delta *  compCost_{Q_{i}{numbatch=1}} 
             + C_{max} \\
            ELSE & \\
             dead&line_{Q_{i}} =   deadline_{Q_{i-1}}  + \\ & \delta * compCost_{Q_{i}{numbatch=1}} 
             \: \forall i= 2 \:to \:n  
    \end{split}
\end{equation}
\end{comment}

\begin{figure*}[tb]
\centering
\fcolorbox{black}{white}{
\def\svgwidth{0.8\textwidth}
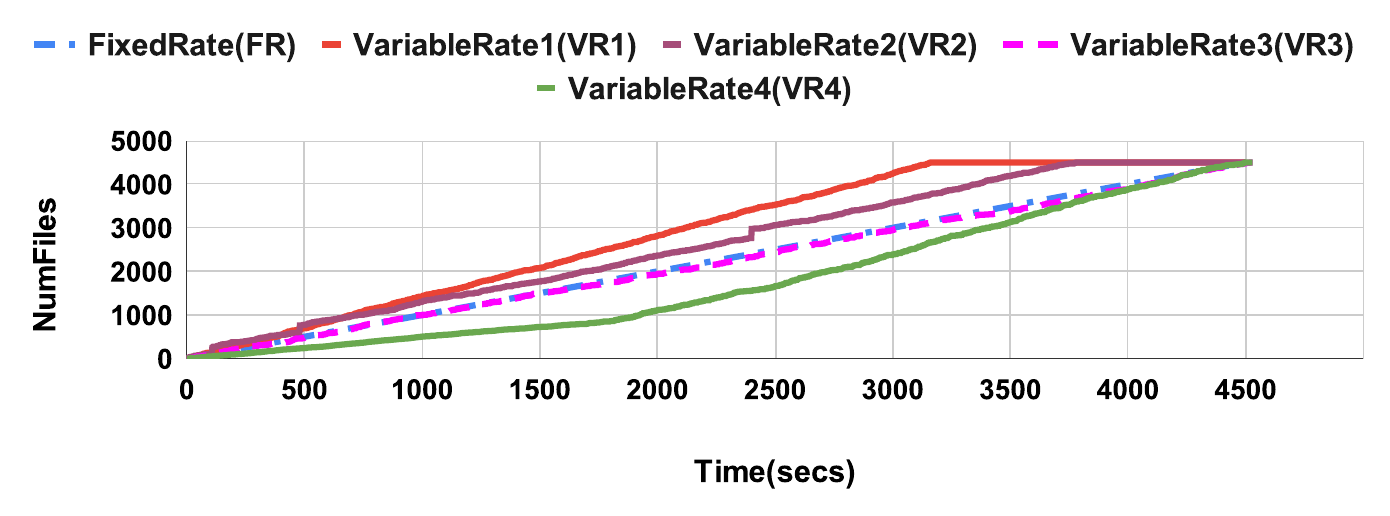}
\newline
\caption{Variable Input Rate For Multi Query Scenario}
\Description{Variable Input Rate For Multi Query Scenario}
\label{fig:variableIpRate}
\end{figure*}

Query submission time was the same as its window start time. Query deadlines were determined with $\delta$ as 1. Scheduling was carried out with LLF, EDF, SJF, and RR strategies. Further, we also ran experiments by setting all queries with the same Window Start time and Window End Time. Different cases were generated by setting $\delta$ to 0.8, 0.6, 0.4, 0.2, and 0.1.

Figure \ref{fig:multiQueryDeadlineFR} gives the results for queries using different strategies with fixed input rate, where the total computation cost incurred by all queries have been normalised with respect to the sum of minimum computation cost of processing all queries in a single batch starting at window end. It can be observed that with EDF and LLF, all the queries were completed within their deadlines except for 0.1D.  EDF and LLF failed for 0.1D as there is no feasible solution with $\delta_{RSF}=50$\%, since the largest deadline for 0.1D was 4594 sec(with window end time is 4500 sec), but the sum of the computation cost for processing the last batch of all the queries was approximately 105 sec. Hence some of the queries could not be completed within their deadlines. 

On the other hand, SJF failed for 0.2D and 0.1D and RR failed for 0.4D, 0.2D and 0.1D as some of the queries could not meet their deadlines.

Thus we conclude that in contrast to earlier work on batching it is essential to consider deadlines while scheduling. Further considering only deadlines without batching may not be beneficial. We ran experiments using EDF and LLF to compare our approach against the traditional scheduling algorithms. In this approach, there is no MinBatchSize determination and the currently available tuples are processed as a batch. Input records are received in terms of files and each file consists of around 9300 tuples. Query to be processed is selected based on earliest deadline or least laxity. 

Both EDF and LLF failed for 0.4D case, while with our method EDF and LLF passed cases upto 0.2D. Also, even for the successful runs, the cost incurred for EDF and LLF were 10 and 7 times more compared to the cost incurred using our approach. Figure \ref{fig:multiQueryDeadlineFR_Trad_EDFLLF} shows the results where the normalised cost with respect to processing in a single batch is more by 15 and 10 times for EDF and LLF respectively. With our approach, the normalised cost for both EDF and LLF was 1.5 times more for $\delta_{RSF}=50$\%. As there is no minimum batch size with the traditional EDF and LLF, and since tuples arrive continuously, queries with the earliest deadline (for EDF scheduling) is processed repeated until completion. Similarly for LLF, queries with the least slack  are processed frequently compared to others. This increases the total number of batches compared to processing with MinBatchSize and thereby leads to higher overall computation cost. Thus our methodology minimises cost while meeting deadline.

\begin{figure}[tb]
\centering
\fcolorbox{black}{white}{
\def\svgwidth{0.45\textwidth}
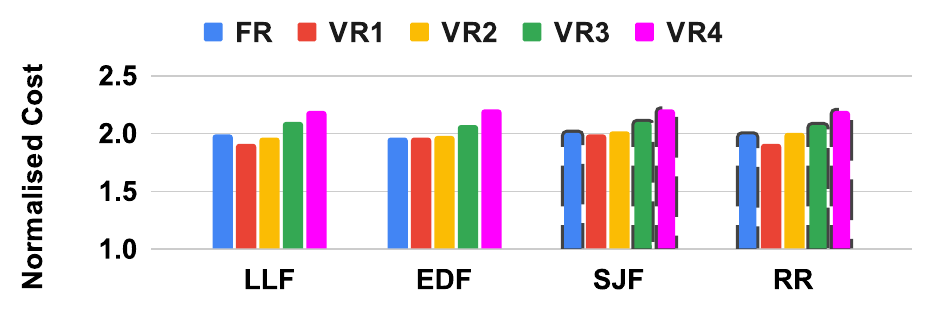}
\newline
\caption{0.1D Cases with $\delta_{RSF}=100$\% }
\Description{0.1D Cases with $\delta_{RSF}=100$\% }
\label{fig:multiQueryDeadlineVR}
\end{figure}

\begin{figure}[tb]
\centering
\fcolorbox{black}{white}{
\def\svgwidth{0.45\textwidth}
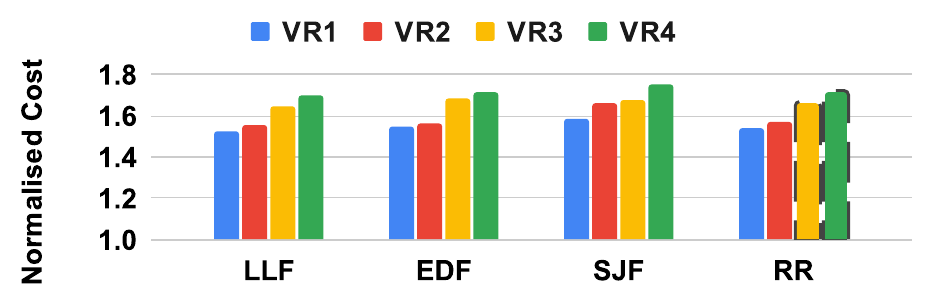}
\newline
\caption{0.2D Cases with $\delta_{RSF}=50$\%}
\Description{0.2D Cases with $\delta_{RSF}=50$\%}
\label{fig:multiQueryDeadlineVR2}
\end{figure}

Then we carried out experiments to assess the impact of increased $\delta_{RSF}$ and variable input rates on the scheduling. We increased $\delta_{RSF}$  to 100\% so that MinBatchSize gets reduced and the number of batches increases. Further to demonstrate the impact of variable input rate on the proposed scheduling schemes, we carried out experiments for 0.1D$\delta_{RSF}=100$\%  with variable input rates. Figure \ref{fig:variableIpRate} shows the fixed and variable input rate profiles. While both VariableRate1 and VariableRate2 are faster compared to the fixed rate, VariableRate2 contains bursty input. Both VariableRate3 and VariableRate4 are slower where the last tuple arrives 20 and 7 seconds delayed, respectively, compared to the fixed rate. Here the the largest deadline is 4594 sec and hence, all the queries have to be processed within 94 seconds of the arrival of the last tuple.

Query scheduling results are shown in Figure \ref{fig:multiQueryDeadlineVR} where the caption denotes fixed (FR) or variable(VR1 to VR4) input rates. The scheduler triggers a batch based on the availability of tuples or on the elapse of time whichever occurs earliest. Both EDF and LLF completed all queries within their deadlines for all cases as in Figure \ref{fig:multiQueryDeadlineVR}. SJF and RR completed all queries for VR1 and VR2. These are runs with faster input rates where all tuples have arrived much ahead of the deadline and hence all queries are completed successfully. For runs with fixed input rate and slower profiles such as VR3 and VR4, both SJF and RR failed as some queries missed their deadlines. 

It can be observed that as $\delta_{RSF}$ is increased from 50\% to 100\%, the normalised computation cost increases from 1.5 (refer Figure \ref{fig:multiQueryDeadlineFR}) to 2 (refer Figure \ref{fig:multiQueryDeadlineVR}). This is due to the increase in the number of batches for $\delta_{RSF}$100\% as compared to $\delta_{RSF}$50\%, thus leading to higher overall computation cost for $\delta_{RSF}$100\%. 

Also, it can be observed that the normalised computation cost is slightly higher for VR3 and VR4 compared to other cases. This is because the number of batches is more for cases with slower input rates. In such cases, even though enough tuples as per MinBatchSize have not been received at the expected time as per the fixed rate profile, the available tuples are processed. For VR4 the total number of batches for processing all the queries is 410 as compared to 329 for FR case. 

Similarly, experiments were carried out with variable input rates for 0.2D with $\delta_{RSF}50$\% and the results are shown in Figure \ref{fig:multiQueryDeadlineVR2}. EDF, LLF and SJF passed all cases while RR failed for VR3 and VR4. For VR3 and VR4 as the actual input rate is slower than the expected rate, initially processing is triggered based on time with the currently available tuples. Further, batches are processed as when enough tuples are ready as per MinBatchSize or based on estimated time. This has significantly reduced the number of tuples to be processed in the last batch after window end. The total number of input files to be processed as part of the last batch after window end  for fixed rate and VR4 is 5620 and 832 respectively. Thus though SJF with fixed rate was not successful, where one query missed its deadline, SJF successfully completed all queries for slower profiles cases, VR3 and VR4, as the total work done in last batch has reduced significantly. But for stringent deadlines like in 0.1D cases, SJF fails for VR3 and VR4. 

To compare our scheme against the existing methodologies, Spark framework do not account for deadlines and in addition, as stated in section 7.1, all queries could not be run simultaneously in Spark streaming due to memory errors. While traditional EDF, LLF may meet deadlines, provided it is feasible, our approach achieves cost reduction due to batching. Also, our approach handles stringent deadline better than traditional EDF and LLF. The results show that our scheduling algorithms complete queries within their deadlines while keeping the overall computation cost not more than $\delta_{RSF}$ fraction compared to the computation cost of processing all tuples in a single batch. Further, the results confirm that EDF and LLF perform better in meeting the query deadlines compared to SJF and RR for both fixed and variable rate input profiles.

\section{Conclusion and Future Work}
\label{sec:concl}

Query processing in high volume systems where data over some period has to be processed and presented at some time point later is a common use case. For such cases, Intermittent Query Processing is ideal as resource consumption can be significantly reduced compared to per-tuple or micro-batch processing. Schemes for scheduling Single and multiple queries under static and dynamic environments have been proposed in this paper. The proposed scheduling strategies  reduce the overall computation cost of a query while honouring its deadline constraints. The results presented in the performance section support the same.   

There are several directions for future work. Our scheduling techniques can be extended for a cluster set up where resources can be added dynamically to complete queries within the deadline. The cost model proposed in this paper can be correlated to the monetary value that would be required for resource allocation. Our current scheduling model runs one batch of one query at a time across all available resources. This can be extended to support the simultaneous execution of different queries on different subsets of nodes in the cluster.

\end{document}